\documentclass[sigconf,nonacm]{aamas}

\usepackage{balance}

\usepackage{hyperref}
\usepackage{subcaption}
\usepackage{array}
\usepackage{multirow}
\usepackage{tablefootnote}

\makeatletter
\gdef\@copyrightpermission{
  \begin{minipage}{0.2\columnwidth}
   \href{https://creativecommons.org/licenses/by/4.0/}{\includegraphics[width=0.90\textwidth]{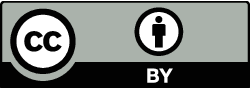}}
  \end{minipage}\hfill
  \begin{minipage}{0.8\columnwidth}
   \href{https://creativecommons.org/licenses/by/4.0/}{This work is licensed under a Creative Commons Attribution International 4.0 License.}
  \end{minipage}
  \vspace{5pt}
}
\makeatother

\setcopyright{ifaamas}
\acmConference[AAMAS '25]{Proc.\@ of the 24th International Conference
on Autonomous Agents and Multiagent Systems (AAMAS 2025)}{May 19 -- 23, 2025}
{Detroit, Michigan, USA}{Y.~Vorobeychik, S.~Das, A.~Nowé  (eds.)}
\copyrightyear{2025}
\acmYear{2025}
\acmDOI{}
\acmPrice{}
\acmISBN{}

\acmSubmissionID{160}

\title[EconoJax]{EconoJax: A Fast \& Scalable Economic Simulation in JAX}

\author{Koen Ponse}
\affiliation{  
  \institution{LIACS, Leiden University}
  \city{Leiden}
  \country{The Netherlands}}
\email{k.ponse@liacs.leidenuniv.nl}

\author{Aske Plaat}
\affiliation{
  \institution{LIACS, Leiden University}
  \city{Leiden}
  \country{The Netherlands}}
\email{aske.plaat@gmail.com}

\author{Niki van Stein}
\affiliation{
  \institution{LIACS, Leiden University}
  \city{Leiden}
  \country{The Netherlands}}
\email{n.van.stein@liacs.leidenuniv.nl}

\author{Thomas M. Moerland}
\affiliation{
  \institution{LIACS, Leiden University}
  \city{Leiden}
  \country{The Netherlands}}
\email{t.m.moerland@liacs.leidenuniv.nl}

\begin{abstract}

Accurate economic simulations often require many experimental runs, particularly when combined with reinforcement learning.
Unfortunately, training reinforcement learning agents in multi-agent economic environments can be slow. 
This paper introduces EconoJax, a fast simulated economy, based on the AI economist \cite{zhengAI2022}.
EconoJax, and its training pipeline, are completely written in JAX.
This allows EconoJax to scale to large population sizes and perform large experiments, while keeping training times within minutes. 
Through experiments with populations of 100 agents, we show how real-world economic behavior emerges through training within 15 minutes, in contrast to previous work that required several days.
We additionally perform experiments in varying sized action spaces to test if some multi-agent methods produce more diverse behavior compared to others. Here, our findings indicate no notable differences in produced behavior with different methods as is sometimes suggested in earlier works. 
To aid further research, we open-source EconoJax on Github at: \url{https://github.com/ponseko/econojax}.

\end{abstract}

\keywords{Reinforcement Learning; Multi-agent; Economics; JAX}

\newcommand{\BibTeX}{\rm B\kern-.05em{\sc i\kern-.025em b}\kern-.08em\TeX}

\begin{document}

%%% The following commands remove the headers in your paper. For final 
%%% papers, these will be inserted during the pagination process.

\pagestyle{fancy}
\fancyhead{}

%%% The next command prints the information defined in the preamble.

\maketitle 

%%%%%%%%%%%%%%%%%%%%%%%%%%%%%%%%%%%%%%%%%%%%%%%%%%%%%%%%%%%%%%%%%%%%%%%%

\section{Introduction}

Reinforcement learning~\cite{suttonbartobook2018} provides an attractive framework for simulating real-world behavior using learned agents in multi-agent environments. Rather than fixing agents to specific strategies, reinforcement learning agents may freely explore the environment and seek economic equilibria without manually implementing certain strategies, which would bias the results.
These agent-based models allow for rich, diverse, and interactive simulations in which economic policy questions may be answered~\cite{zhengAI2022, trottBuil2021}.

However, reinforcement learning is known to be sample inefficient~\cite{kaiserMode2024}, delivering significantly less information per datapoint compared to practices such as supervised or self-supervised learning. Furthermore, in online reinforcement learning, data collection occurs concurrently with the learning process, causing a further increase in computational requirements. 
Consequently, reinforcement learning agents commonly require days of training, even in simplified, unrealistic simulations.

Over the past decade, we have seen significant performance improvements in deep learning by leveraging efficient GPU's~\cite{krizhevskyImag2012a}. Although this partly aids deep reinforcement learning agents in their training, real-time environment interactions needed for online learning remain costly.
In the last few years, however, tools have been built that allow for faster environment interaction by taking advantage of GPU's.
Of particular interest are WarpDrive~\cite{lanWarp2021}, and JAX~\cite{bradburyJAX2018}, which both provide methods of writing environments to run on GPU accelerated hardware. 
WarpDrive, however, requires the developer to write specialized CUDA code, which may be difficult to master.
In contrast, JAX code is written in regular Python code, albeit with some conditions to the coding style.
Importantly, we can also write the reinforcement learning agent code in JAX and unify all code (environment and agent) under one framework.
This code can then be compiled by JAX as a single training loop pipeline that operates fully on accelerated hardware, yielding substantial performance improvements, sometimes over a 1000 times ~\cite{luDisc2022}.

\begin{figure*}[ht] \label{fig:overview}
    \centering
    \includegraphics[width=0.95\linewidth]{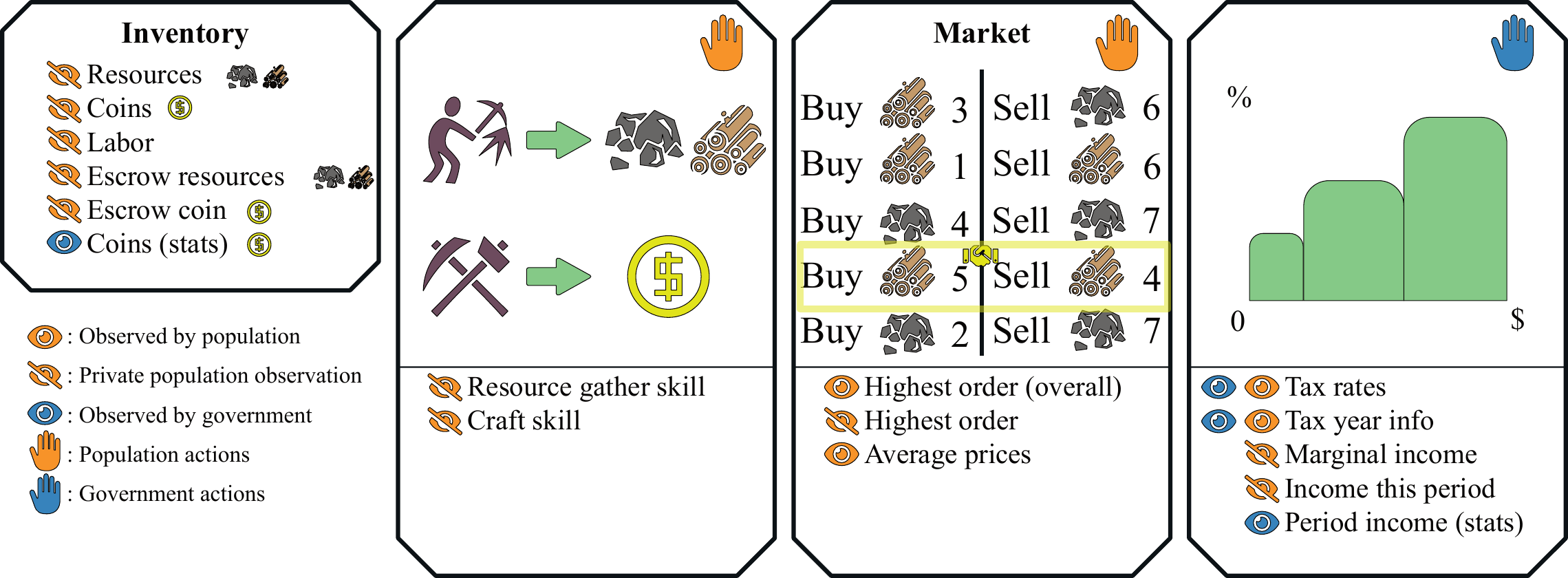}
    \caption{Overview of EconoJax and its different components, actions- and observation spaces. Population agents act via gathering resources, converting these into coin, and trading on the marketplace. The government agent acts by setting tax rates at specific intervals. The escrow inventories are used as a temporary inventory for items that are used in active market orders. Observations marked with "stats", indicate that the government does not observe the state for each individual agent, but rather the mean, standard deviation, and median of the population.}
    \label{fig:world-overview}
    \Description{EconoJax overview of action and observation space}
\end{figure*}

In recent years, the scientific community has made a push to reintroduce existing environments in JAX to address the sample efficiency problem~\cite{matthewsCraf2024a, gymnax2022github, pignatelliNAVI2024, rutherfordJaxM2023, luDisc2022, bonnetJuma2024, brax2021github, lu2023structured, frey2023jaxlob}.
However, a similar initiative has not yet emerged for economic agent-based modeling environments. Such an effort could yield significant benefits, as it would allow for richer, more realistic environments, while keeping agent training feasible.

In this work, we introduce EconoJax, a multi-agent economic environment simulating a simplified economy modeled after the AI economist ~\cite{zhengAI2022}, written in JAX.
EconoJax is a vector-based environment with similar functionalities to the AI economist. Population agents gather resources and convert these into coins to maximize their individual utility. The agents interact by trading resources in exchange for coins with other agents. Simultaneously, a government agent sets tax rates at predefined intervals to maximize equality among agents while preserving productivity in the economy. 

We experiment with population sizes of 100 agents, which requires roughly 13 minutes of training, and highlight real-world economic behavior emerging in EconoJax, such as realistic \textit{tax schedules} and the \textit{equality-productivity-tradeoff}, without manually implementing actions for agents.
In comparison, such experiments took many days of training in the original AI economist framework due to a more complex state representation.
Due to its increased performance, while still allowing for real-world economic behavior to occur, EconoJax provides a foundation for economic modeling researchers to build richer and more realistic simulations.

EconoJax can also flexibly scale to larger population sizes and action spaces. By increasing the number of available resources, agents have the option to specialize in more diverse behavior to optimize their own utility. This makes EconoJax an interesting testbed for multi-agent RL research. We explore the effect of various multi-agent methods on agent behavior, noting that we observe a roughly equal spread in behavior for agents that are trained individually, compared to agents that share all their network parameters.

The contributions of this paper are as follows:
\begin{itemize}
    \item We present EconoJax, a scalable economic simulation written in JAX. EconoJax, inspired by the AI economist, provides a simplified and substantially faster interface for reinforcement learning training, highlighting how economic simulations may benefit from JAX.
    \item We conduct experiments to demonstrate how real-world economic behavior emerges in EconoJax, similar to the AI economist.
    \item We use EconoJax for additional experiments comparing different multi-agent learning methods in varying sizes of action spaces. 
    \item We release the source code for EconoJax, along with the reinforcement learning agent code. This lowers the bar to extend or experiment with EconoJax and may inspire further research in economic simulations and multi-agent reinforcement learning.
    
\end{itemize}

The remainder of this paper is organized in the following manner: In the next section, Section \ref{sec:preliminaries}, we give a brief overview of reinforcement learning, an introduction to JAX, and discuss related work.
Section \ref{sec:econojax} details of EconoJax, including its agents and their respective action and observation spaces. 
Subsequently, Section \ref{sec:experiments} displays the outcomes of experiments conducted in EconoJax involving 100 agents.
Next, in Section \ref{sec:multi-agent}, we experiment with different multi-agent methods to showcase their effect on produced policies in varying action space sizes.
In Section \ref{sec:discussion} we discuss limitations and future work.
Finally, our conclusions are summarized in Section \ref{sec:conclusion}.

\section{Preliminaries} \label{sec:preliminaries}

\subsection*{Reinforcement Learning}

Reinforcement Learning is a machine learning method for finding policies that attempt to maximize some reward by interacting with an environment~\cite{suttonbartobook2018}. Such environments are typically modeled as a Markov Decision Process (MDP)~\cite{bellmanMark1957}, defined as a tuple ($\mathcal{S}$, $\mathcal{A}, p, \mathcal{R}, \gamma$).
Here, $\mathcal{S}$ is the set of all states, $\mathcal{A}$ the set all possible actions available to the agents, $p$ is the transition function which maps states and actions into a probability distribution of next states $p(s'|s,a)$ : $\mathcal{S} x \mathcal{A} x \mathcal{S} \mapsto [0,1]$, $\mathcal{R}$ is the reward function, mapping transitions into rewards $r : \mathcal{S} x \mathcal{A} x \mathcal{S} \mapsto \mathbb{R}$, and $\gamma$ is the discount factor that governs the effect of future rewards on the agent. 

More informally, reinforcement learning provides us with a learning framework consisting of an environment and an agent (or multiple agents). The environment starts in a state $s \in \mathcal{S}$ that is (partially) observable to the agent(s). The agents then produce an action $a$ from the available set of actions $\mathcal{A}$, which triggers the transition of the environment into a new state. The agents are then also provided a \textit{reward} $r$, which is a measure of how "good" the new state is. This process is repeated with the goal of finding an optimal \textit{policy} for the agent(s) to follow in the environment. 

\subsection*{JAX}

JAX is a python library for accelerator oriënted programming~\cite{bradburyJAX2018}, providing convenient function transformations, just-in-time compilation, and a NumPy interface. 
Just-in-time-compilation support allows JAX functions to be compiled with XLA~\cite{xla_citation}, such that these functions may efficiently run on platforms such as CPU's, GPU's or TPU's.
Furthermore, JAX offers transformations that can easily vectorize, parallelize, or differentiate Python JAX functions. 
The built-in functionality to differentiate functions (autograd) allows JAX to be used as a deep learning framework, similar to PyTorch and TensorFlow. As such, a large ecosystem of neural network libraries have been built around JAX~\cite{kidger2021equinox, haiku2020github, jraph2020github, flax2020github}.

JAX allows all the aforementioned functionality on functions which are written in plain Python. In turn, these typically difficult practices (e.g. running code on a GPU, vectorization, or parallelization), are now significantly more accessible and can dramatically speed up performance in machine learning workflows. Importantly, it allows us to write reinforcement learning environments in plain Python that can easily run on the GPU.

Unfortunately, while JAX functions are written in plain Python, they have some restrictions. The most notable of these is that JAX transformed functions must be \textit{functionally pure}, disallowing any side-effects.
This feels unintuitive to many developers that write object-oriented code and rely heavily on changing an objects \textit{state}, which is not allowed in functionally pure code.
JAX does provide a NumPy-style interface such that most JAX restrictions can be addressed via an API that is familiar to most Python developers.
However, when starting with JAX, it is clear that a different style of coding is required, which can feel restrictive.

The restrictions imposed by JAX often do not set hard limits on what is possible. This paper exemplifies this by converting the AI economist into a similar environment written in JAX. In the next section, we will list more libraries that have created new, or recreated existing environments in JAX, and by extension, allow for much faster training times.

\subsection*{Related Work}

As JAX has increased in popularity, more environments have been reimplemented in the framework.
As such, we have JAX based environment suites featuring classic environments such as Gymnax and Jumanji~\cite{gymnax2022github, bonnetJuma2024}, as well as suites that focus on multi-agent reinforcement learning~\cite{rutherfordJaxM2023, pax, dekock2023mava}, partial observability~\cite{lu2023structured}, Physics engines~\cite{brax2021github}, and board games~\cite{koyamada2023pgx}.
Furthermore, larger, single environments, rather than suites, are also built in JAX, such as Craftax~\cite{matthewsCraf2024a} and Navix~\cite{pignatelliNAVI2024} (MiniGrid).

Agent-based model environments coupled with multi-agent reinforcement learning training could provide interesting insights into (hypothetical) policies and their effect on society.
The AI economist~\cite{zhengAI2022} demonstrated this by highlighting how reinforcement learning could be used to study the effects of different tax schedules in a simplified economy. They further showed that this would then allow for these tax schedules to be learned as well.
This framework was used in a similar fashion to study US state policies for combatting COVID-19~\cite{trottBuil2021}.
More complex economic simulations have recently been built as well, modeling governments, banks, households and firms~\cite{dwarakanatABID2024}.
Other works have studied, using reinforcement learning, how fictitious regions can best negotiate agreements to combat climate change~\cite{mila-competition}.

Recently, an agent-based modeling framework was built in JAX to model the foraging behavior of large populations of organisms~\cite{chaturvediFora2024}. However, to the best of our knowledge, no simulator for an economy has yet been released in JAX.

\begin{figure*}[ht]
\centering
    \subcaptionbox{Productivty}{\includegraphics[width=0.33\textwidth]{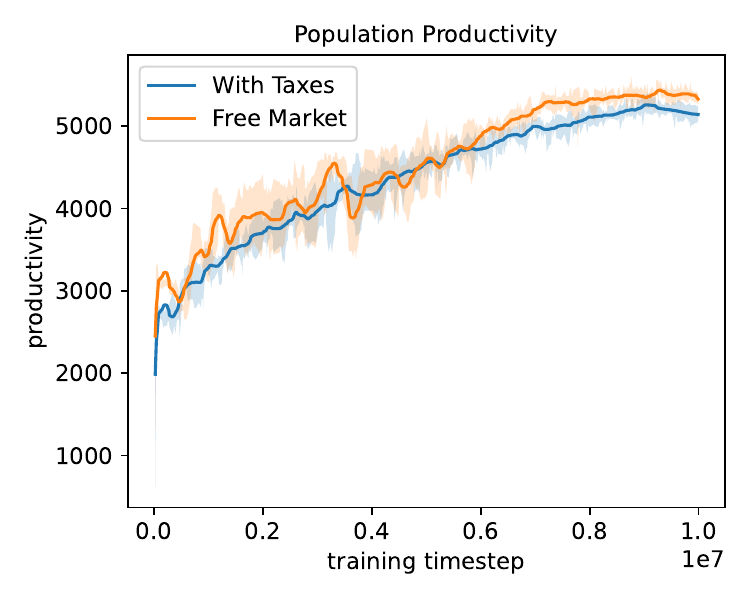}}%
    \hfill
    \subcaptionbox{Equality: (1 - gini)}{\includegraphics[width=0.33\textwidth]{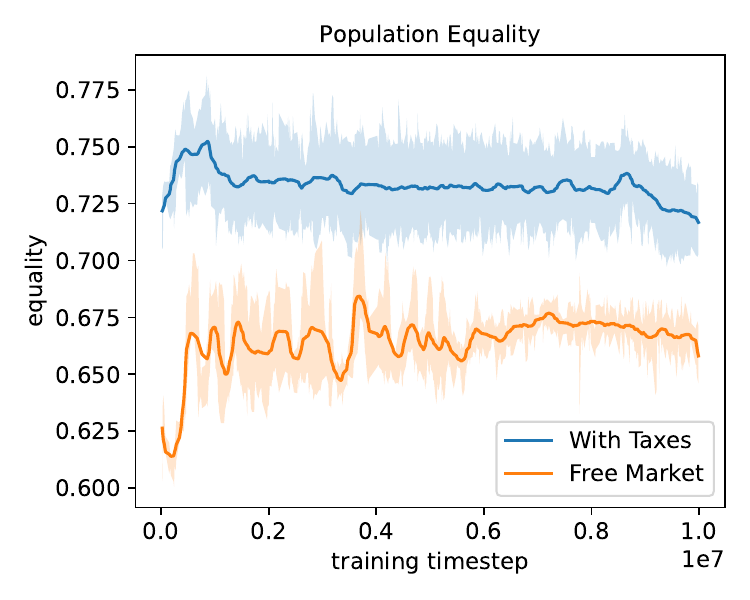}}%
    \subcaptionbox{Government Utility}{\includegraphics[width=0.33\textwidth]{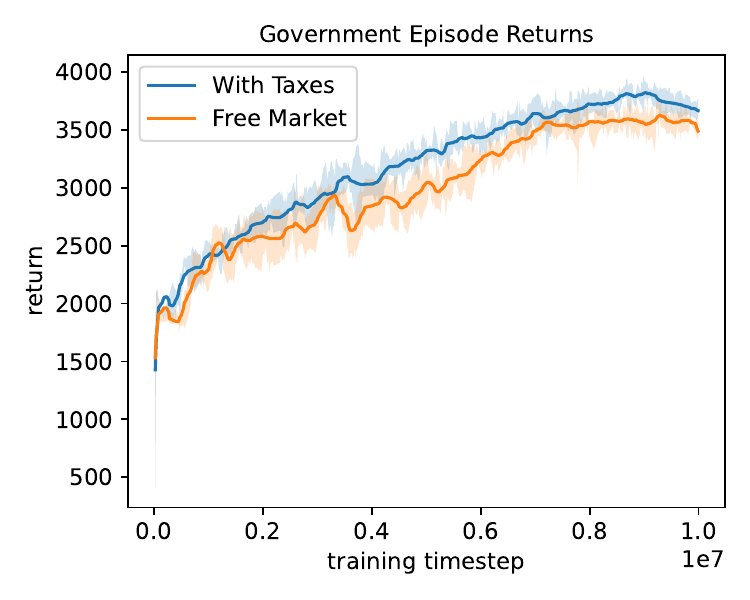}}%
\Description{Productivity and Equality curves over training Productivity and Equality curves over training}
\caption{Productivity and equality measured at the end of an episode during training. The shaded areas represent the standard deviation of 15 training runs, in each of which 10 different environments ran in parallel. We see productivity increases over time due to the population agents learning. Introducing taxes manages to significantly improve equality in the population, but trades this for a small bit of productivity. Government utility (c) is measured as productivity $\cdot$ equality, weighted equally. Near the end of training, the government found appropriate tax rates such that the population has close to optimal productivity, while achieving higher equality.}
\label{fig:equality-and-productivity}%
\end{figure*}

\section{EconoJax} \label{sec:econojax}

In this section, we explain the design of EconoJax, an economic simulation modeled after the AI economist~\cite{zhengAI2022}, created in JAX. The EconoJax source code, along with the experiment code and PPO~\cite{ppo_paper} agent, is available on GitHub: \url{https://github.com/ponseko/econojax}, allowing researchers to modify EconoJax for their own research.

\subsection*{Overall design}

EconoJax is a simulation environment for a simple economy modeled after the AI economist framework~\cite{zhengAI2022}. Similarly to the environments created in the AI economist, EconoJax closely follows the Gymnasium API~\cite{towersGymn2024} (with modification inspired by Gymnax~\cite{gymnax2022github}) and is thus formally an MDP, typically used for reinforcement learning research.

Like the AI economist, EconoJax features two types of agents -- a set of $n$ \textbf{population agents}, and a single \textbf{government} agent. The \textbf{population agents} are able to gather resources, convert these resources into coins, and trade with each other. Their goal is to maximize their individual utility, a function of coin and performed labor.
Meanwhile, the \textbf{government agent} is able to set tax rates such that taxes can be collected and redistributed at set intervals. Its goal is to maximize the equality and productivity of the population.
Unlike the AI economist, EconoJax does not feature a 2D grid world in which the \textbf{population agents} move. Instead, agents do not move and alternatively receive a simplified vector representation (1D) as input. Such a vector representation allows for faster training without compromising realism. In Section \ref{sec:experiments}, we highlight how real-world economic behavior still emerges in EconoJax.

In the following two sections, we dive deeper into the actions and reward functions of both types of agents, along with the built-in parameters of EconoJax to alter the environment.
A complete overview of the action and observation space of both agents is shown in Figure \ref{fig:world-overview}.

\subsection*{Population Agents}

Population agents represent economic actors that optimize to their own happiness, represented as utility. This utility grows by obtaining \textit{coins} and decreases by performing \textit{labor}. 
Similarly to a real economy, agents have various actions of obtaining coins, all of which provide a set amount of labor when performed. 
Population agents can perform one action per timestep, which may be collecting resources, converting these resources into coins, or interact with other agents by trading resources in exchange for coins.

\paragraph{Reward function} In the default EconoJax implementation, population agents are tasked with maximizing their individual discounted utility. Similar to the AI economist, we model utility as isoelastic utility~\cite{arrow1971theory}, minus their obtained labor $L$:
\begin{equation}
    u_{i,t} = \frac{C_{i,t}^{1-\eta} - 1}{1 - \eta} - L_{i,t}
\end{equation}
Here, $C_{i,t}$ is the total coin in the inventory of agent $i$ at timestep $t$ and $\eta$ ($\eta \geq 0, \eta \neq 1$) is a constant determining the concavity of the utility function -- the isoelastic utility function will increase at a slower rate the more coin an agent already has obtained, and a high $\eta$ increases this effect. We set $\eta$ as 0.27 in all our experiments, mirroring the AI economist.
The final agent reward on a specific timestep is given by the change in utility at every timestep:
\begin{equation} \label{eq:rew-diff}
    r_{i,t} = u_{i,t+1} - u_{i,t}
\end{equation}

\paragraph*{Collecting resources} 
An EconoJax environment contains resources that agents can collect. The amount of different types of resources can be set to any positive number such that the agents may specialize in different types, allowing users to flexibly scale the economy. 
This is an abstraction of the \textit{wood} and \textit{stone} resources in the AI economist.
The amount of resources obtained by an agent when gathering resources is determined by an agents' skill level $SG_{i,j}$, where $j$ is the resource id.
The skill levels are set at environment initialization (uniform between 0 and 1 by default) and fixed throughout the episode.
When attempting to gather a resource, the exact amount of resources gathered is then given by $\lfloor SG_{i,j} + \rho \rfloor$, where $\rho$ is a random number between $0.0$ and $1.1$, chosen uniformly.
As such, low-skilled agents may still successfully gather a resource if lucky and highly-skilled agents may gather multiple resources in a single action.
Gathered resources are stored in the agents personal inventory for further use by one of the other actions.

\paragraph*{Crafting} 
When agents have obtained a set amount of required resources, they can convert these resources into coins.
The amount of required resources is set by environment parameters $R_\kappa$ and $R_d$, where $R_\kappa$ describes the number of resources required per resource, and $R_d$ describes the amount of different resources required. For instance, an environment initialized with 5 different resources and $R_\kappa=2$ and $R_d=3$, means an agent is required to have 2 of each of 3 resources in their inventory before they can craft.
In case multiple different resources may be used for crafting, the resources used are prioritized by the amount the agent has in its inventory. 
The amount of coins obtained by crafting is determined by the skill level of an agent $SC_{i}$, scaled by a constant. $SC_{i}$ is initialized at environment initialization (uniform between 0 and 1 by default) and fixed throughout the episode.

\paragraph*{Trading} 
Rather than crafting coin, agents are also able to sell their resources to other agents in exchange for coins, allowing for interaction between agents.
To facilitate trading, agents can place buy or sell orders on a marketplace. Orders are active for a certain amount of time, and at each timestep all current orders are evaluated.
In case there exist buy orders with a value higher or equal to a sell order, these orders are matched -- resources and coin are exchanged.
In case orders could be matched to multiple other orders, high-value orders (high bids / low asks) are evaluated first. Ties are then first broken by the age of the order and lastly at random.
In order to facilitate trade, coins and resources that are currently used for orders are temporarily placed in agents \textit{escrow}, rather than their usual inventory. 

\vspace{1em}
Optionally, an additional \textit{NO-OP} action may be added to the action space of population agents to allow them to skip performing an action and thus collect no labor.

\subsection*{Government Agent (Planner)}

The government agent sets tax rates in fixed tax brackets with the goal of maximizing equality and productivity. These two measures are seen as important optimization goals for taxation (and the redistribution of those taxes)~\cite{nationsunIneq2013}, but this objective may be altered in EconoJax. 
Tax rates are set every period (default: 100 timesteps). At the end of each period, tax is collected on the earnings from the previous period. The collected taxes are then redistributed uniformly among agents, and new tax rates may be set. 

\paragraph{Reward function} As stated, the government utility (social welfare) is determined by the productivity and equality in the environment:
\begin{equation}
    u_{g} = \text{Eq}(C_{t}) \cdot \text{Prod}(C_{t}),
\end{equation}
\begin{equation}
    \text{Eq}(C_{t}) = \omega(1 - \text{gini}(C_{t})) + 1 + \omega, 0 \leq \omega \leq 1.
\end{equation}
Here, $\omega$ is the equality weight, determining how much equality weights on social welfare (defaults to 1, weighing productivity and equality equally). 
$\text{prod}(C_{t})$ is the total amount of coins in the environment:
\begin{equation}
    \text{Prod}(C_{t}) = \sum_i{C_{i,t}},
\end{equation}
and $\text{gini}(C_t)$ is the gini index~\cite{gini1912variability}, a well-known measure of equality. 
Similarly to the population agents, the final reward for the government agent is the difference in $u_g$ between timesteps, as in Equation~\ref{eq:rew-diff}.

\paragraph{Setting tax rates}
We divide an environment episode into periods of a fixed number of time steps. At the start of every period, the government can set new tax rates for the upcoming period. 
In the same timestep, taxes of the previous period are collected and redistributed uniformly among all agents.
Tax rates may be set at 20 intervals of 5\% in predetermined tax brackets (0\%, 5\%, 10\% ... 100\%). 
For example, given tax brackets \textit{[50;100]}, the government can set three tax rates for each of the intervals: $[0,50), [50,100), [100,\infty]$, on 20 different levels. 
Given three tax rates such as (10\%, 30\%, 50\%), an agent that earned 130 coin in the previous period will pay $0.10 \cdot 50 + 0.30 \cdot 50 + 0.50 \cdot 30 = 35$ coins in taxes. A fraction of these taxes ($1/n$) will immediately be returned to the agent due to the redistribution.

\subsection*{Action masking}

We use action masking to prevent invalid actions. For example, an agent can only sell a resource if it has at least one unit of that resource in its inventory. Action masking assigns a zero-probability to any not-available action and thereby improves training times as agents cannot perform pointless actions.
Although the actions of the government only have an effect at some timesteps, it does observe all intermediate steps, enabling models that use historical data to benefit from these observations.

\begin{table}[hb!]
    \centering
    \caption{Training times in hours for both EconoJax (averaged over 30 runs) and the AI economist (averaged over 6 runs).
    For the AI economist, we used the available RLlib implementation, using 32 CPU threads.
    For this comparison, we have had to use significantly different hardware. Therefor, the two environments cannot be directly compared. In turn, this table only gives a rough indication of the performance benefits of EconoJax.
    In a same sized population, EconoJax trains its agents over 2.400 times faster compared to the AI economist. 
    This is in large part due to the simplified state representation of EconoJax and the GPU based training powered by JAX.}
    \begin{tabular}{l|cc}
        \hline
                                & \textbf{AI Economist (CPU)} & \textbf{EconoJax (GPU)}     \\
                                & pop. size: 4  & pop. size: 4 \\
        \hline
        
        \rule{0pt}{5mm} Until convergence\footnotemark      &  168.28     &  0.07                 \\
        \rule{0pt}{5mm} Per million steps &  1.12 &  0.01                 \\
        \hline
    \end{tabular}
    
    \label{tab:ai_economist_vs_econojax}
\end{table}

\footnotetext{We trained the AI economist agents for 150 million steps. By then, results had not fully converged yet, but training for longer was hindered by large memory requirements. EconoJax was trained for 5 million timesteps (10 million in experiments with population sizes of 100).}

\section{Experiments} \label{sec:experiments}

In this section, we discuss results of experiments similar to those of the AI economist and demonstrate how real-world economic behavior emerges in EconoJax such as \textit{specialization}, the \textit{productivity-equality tradeoff}, and realistic \textit{tax schedules} and \textit{market prices}. We compare the free market setting (no government agent) to an environment with the government agent enabled, attempting to maximize equality and productivity, weighing both of those factors equally.

Training EconoJax environments with a population of 100 agents took only about 13 minutes on a single \texttt{NVIDIA A100} GPU. 
In Table~\ref{tab:ai_economist_vs_econojax}, we compare EconoJax to the AI Economist by training with only 4 agents. In this setting, EconoJax finishes training in roughly 4 minutes. 
In contrast, reproducing the AI economist results for the same sized population, 
using the available RLlib~\cite{liangRLli2018a} implementation, took a full week to complete roughly 150 million training steps on a \texttt{16 Intel Xeon E5-2630v3 cores @ 2.40GHz (32 threads)}, before often running out of memory, disallowing us to complete the full 1 billion training steps required. 
Due to the difference in hardware and observation space (2D vs 1D observations), this is not a fair direct comparison. 
However, this \textbf{$>$2400$\times$ performance improvement} in a population of the same size, does provide a clear indication of the substantial decrease in the computational requirements of EconoJax.
These reductions allow researchers to experiment with less hardware and enable future expansions of EconoJax for enhanced realism while still keeping training times feasible.

\subsection*{Experimental details}

We have experimented in EconoJax with population sizes of 100, far exceeding the population sizes of 4 and 10 experimented with in the AI economist. 
We trained all our agents using PPO~\cite{ppo_paper}, with network sizes of two layers with 128 nodes per layer and tangent activation functions. In these experiments, all population agents shared weights for both the policy and value network.
Both population and government agents are able to train simultaneously, in contrast to the two-stage training process required in the AI economist. 
The results of our experiments are obtained by averaging over 10 different environment seeds during training. Additionally, we repeated training 5 times for 3 different (but similar) population skill distributions and highlight the standard deviation over these 15 runs in shaded areas.
The population skill levels are initialized via a Pareto distribution with added random noise. This ensures that most agents are at least reasonably skilled in one action.

The complete list of hyperparameters of the agents, as well as the environment parameters used in these experiments, are listed in Table \ref{tab:(hyper)parameters}.

\begin{table}[ht]
    \centering
    \caption{Hyperparameters for our PPO agents (top), and EconoJax environment parameters used for our experiments (bottom) in Section \ref{sec:experiments}. Both government and population agents used the same hyperparameters during training. Entropy coefficient is linearly annealed over 90\% of training steps, whereas learning rate annealling happens over the full training duration. 
    Agent skill levels in our experiments, are set such that most agents are at least reasonably skilled at one action, via a pareto distribution. Random normal noise is then added to the skills levels for a less "static" population.}
    \begin{tabular}{lr}
        \toprule
        \textbf{Hyperparameter} & \textbf{Value} \\
        \midrule
        Training Timesteps & 1e7 \\
        Learning Rate & 0.0005 (annealed) \\
        Discount Factor ($\gamma$) & 0.999 \\
        Clip parameter & 0.2 \\
        Entropy Coefficient & 0.1 (annealed) \\
        Value Function Coefficient & 0.25 \\
        Value Clip Parameter & 10.0 \\ 
        GAE $\lambda$ & 0.95 \\
        Rollout Length & 150 \\
        Number of Epochs, Minibatches & 6, 6 \\
        Number of Layers, Nodes & 2, 128 \\
        \midrule
        \textbf{Parameter} & \textbf{Description} \\
        \midrule
        Population size & 100 \\
        Number of resources & 2 \\
        Episode length (in steps) & 1000 \\
        Tax Period Length (in steps) & 100 \\
        Allow NOOP Action & True \\
        Starting Coin & 15 \\ 
        Trade Order Expire time & 30 \\
        Possible trade prices & [2, 4, 6, 8, 10] \\
        Maximum active orders per agent & 15 \\
        Craft Resources Required ($R_\kappa$) & 2 \\ 
        Craft Diff. Resources Required ($R_d$) & 2 \\
        Labor Cost Crafting & 1 \\
        Labor Cost Gathering & 1 \\
        Labor Cost Trading & 0.05 \\
        isoelastic utility $\eta$ & 0.27 \\
        equality weight ($\omega$) & 1 \\
        \bottomrule
    \end{tabular}
    \label{tab:(hyper)parameters}
\end{table}

\subsection*{Results}

Figure \ref{fig:equality-and-productivity} visualizes the productivity and equality during training of 10 million steps (100.000 tax periods). Introducing taxes increases equality among the population, but this comes at the cost of productivity. 
This is the result of the \textit{productivity-equality trade-off} -- taxes cause some agents to perform less total labor as they no longer experience an increase in utility. Nevertheless, the government has managed to find tax schedules that overall cause minimal decrease in productivity (Figure \ref{fig:equality-and-productivity}a). As a result, the government manages to improve on its goal compared to a free market, as shown in Figure~\ref{fig:equality-and-productivity}c.
Note that, because the population and government agents train simultaneously, an increase in equality does not produce a clear decrease in productivity. This is because the population agents are becoming more efficient during training, causing a natural increase in productivity regardless of taxation.

\begin{figure}[ht]
    \centering
    \includegraphics[width=0.45\textwidth]{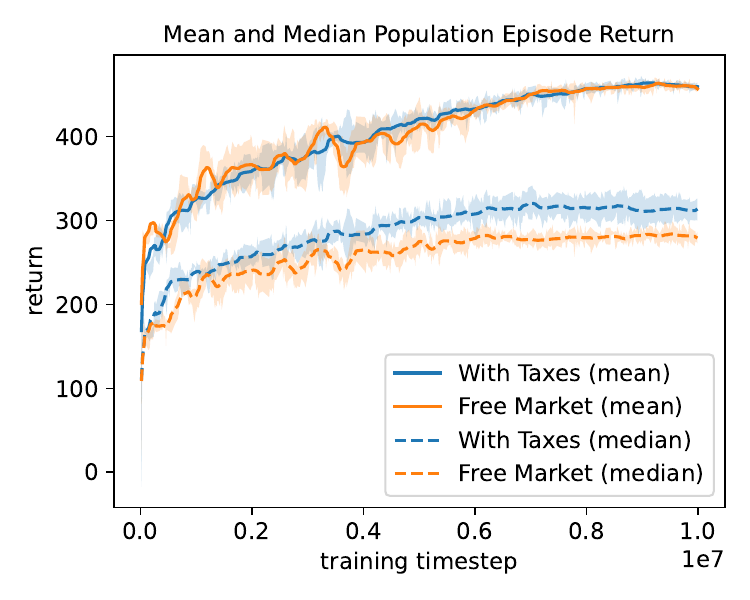}
    \Description{Population mean and median rewards over training}
    \caption{Population mean and median episode returns during training. The shaded areas represent the standard deviation of 15 training runs, in each of which 10 different environments ran in parallel. At the end of training, the utility of the population is roughly equal in both economic systems. However, the median of the population is substantially higher when taxes are introduced -- indicating that a larger share of the population prefers the systems with taxes in place.}
\label{fig:pop-rewards}%
\end{figure}

On average, at the end of training, population agents achieve similar levels of utility in both economic systems, as depicted in Figure \ref{fig:pop-rewards}. However, the median utility, which could be considered a more accurate indicator of overall economic happiness, improves notably with the implementation of taxes, reflecting the increase in equality shown in Figure \ref{fig:equality-and-productivity}b.

\begin{figure}[ht]
    \centering
    \includegraphics[width=0.8\linewidth]{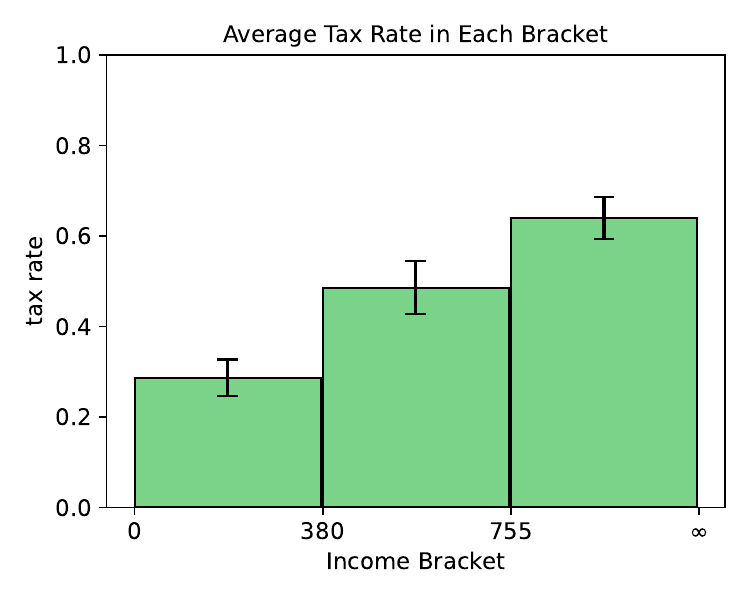}
    \caption{Average produced tax brackets produced by the government agent. The tax rates are the average tax rates produced for three different population groups with different (but similar) skill distributions. For each skill distribution, we retrained agents on 5 different seeds and evaluate over 15 different environment seeds each. The error bars indicate the standard error. We observe a progressive tax system, which is common in many countries around the globe, and was not naturally produced in the AI economist.}
    \label{fig:tax-rates}
    \Description{Tax rates produced by the government agent after training}
\end{figure}

As stated above, default EconoJax uses fixed tax brackets thresholds, allowing the government to set rates within these brackets. For these experiments, we used three brackets that resemble the Dutch tax brackets for 2025, scaled down by a factor 100.
The resulting tax schedules, proposed by the government agent, are shown in Figure \ref{fig:tax-rates}.
Interestingly, the EconoJax government agent produces a progressive tax schedule, common in many countries around the world. We often observed this pattern in our experiments, including those with different population sizes and skill distributions. 
However, while less common, degressive tax systems or \textit{"M"} shaped systems were sometimes also generated in different runs. In those runs, the government found an optima by using the tax system to push the population into working. 
We should note that all of these results rely heavily on the choices made in the environment, which we will discuss further in Section \ref{sec:discussion}.

\begin{figure}
    \centering
    \includegraphics[width=\linewidth]{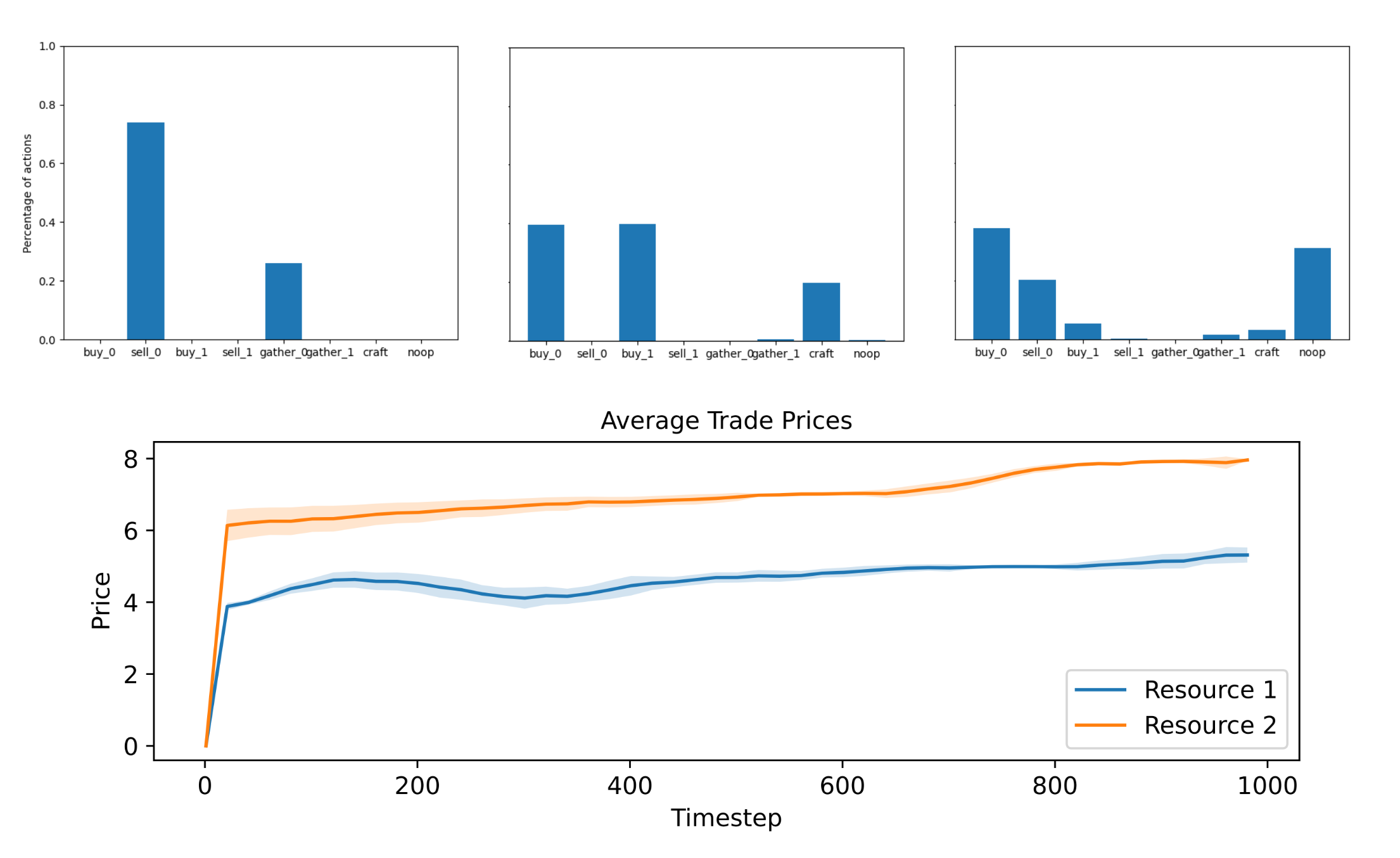}
    \caption{Top: three action distributions of different agents in the same simulation, showcasing specialization in different tasks, that are beneficial due to their skill level. These three agents gather and sell, buy and craft, or only trade / perform no labor.
    The bottom figure shows prices in an environment episode, averaged over 15 environments. These environments were initialized with more agents (roughly 2x) proficient in gathering \texttt{resource 1}. As we would expect, the average trading price for this resource is noticeably lower.}
    \label{fig:action-dist-and-prices}
    \Description{Action distribution examples of three agents and average price development over time.}
\end{figure}

We further observed that, after training, the government produced steady tax rates, not changing the rates within a single environment episode.
This is reasonable, given that the economy remained stable -- during the simulation, there was neither an influx nor an outflow of people, and skill levels did not change.
Similarly, the population agents also stabilized into a specialized role. Focusing on different tasks like buying and crafting, or gathering and selling.
We also observed a realistic market signal in the average sale prices for the resources. That is, resources for which a smaller part of the population had a high skill level became more expensive on average. We show this, along with specialized action distributions in Figure \ref{fig:action-dist-and-prices}.

\section{Multi-Agent Methods} \label{sec:multi-agent}

\begin{figure}[ht]
\centering
    \subcaptionbox{Agent Returns (r=4)}{\includegraphics[width=0.40\textwidth]{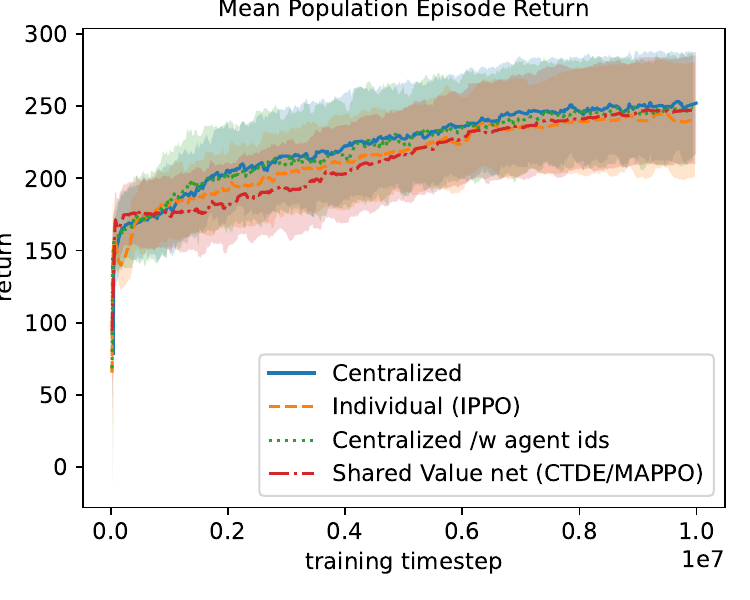}}%
    \hfill
    \subcaptionbox{Agent Returns (r=12)}{\includegraphics[width=0.40\textwidth]{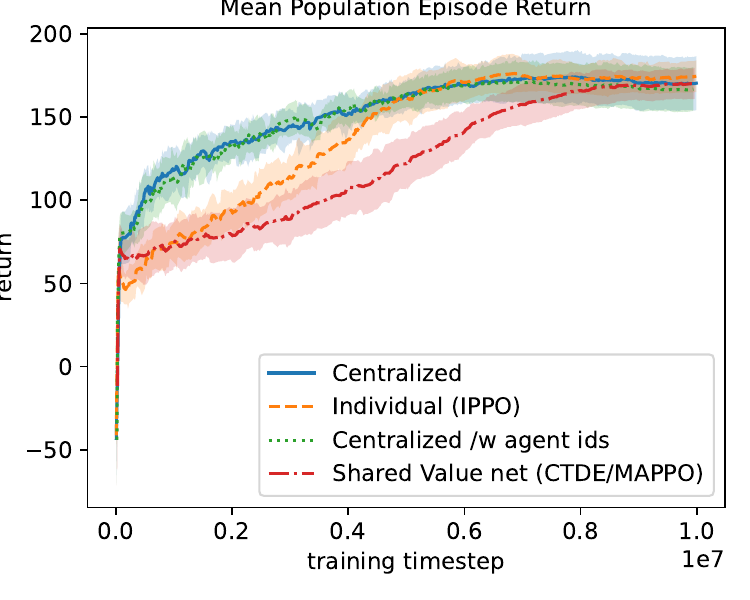}}%
\Description{\ldots}
\caption{Agent returns with various standard multi-agent practices for different amount of resources in EconoJax. Higher resources create a larger action space and more possibilities for agents to produce diverse behavior. Our results indicate that centralized training produces roughly the same behavior compared to training individual networks.}
\label{fig:multi-agent-comparison}%
\end{figure}

In this section, we perform additional experiments in EconoJax with an increasing number of resources. By increasing the number of resources, the possible variety in agent behavior grows, which allows us to test the effect of different multi-agent methods in terms of their diversity.

For these experiments, we inspect three common methods for multi-agent RL: 1) centralized training, sharing all network parameters among all agents. Agents will behave differently due to the different private observations given to the network for each agent; 2) Independent training, each agent training their own set of network parameters; 3) In actor-critic methods, where RL agents train a value- and policy network separately, we can mix both prior methods by training individual policy networks and a central value network. Methods utilizing this technique are often referred to as \textit{centralized training, decentralized execution (CTDE)} methods. For this last method, we only consider the naive version in which the value network observes the same type of observation as the policy network (with larger batches due to the network observing the observation for each agent). Although there are more sophisticated CTDE methods~\cite{kubaTrus2022} that may obtain higher performance, EconoJax agents are tasked with mimicking the real world and enhancing their value network observation arguably degrades their realism.

We used PPO~\cite{ppo_paper} as the learning algorithm for all three of our methods. The independent varient is then sometimes referred to as iPPO, and the sharing of the value network as MAPPO~\cite{yuSurp2022}.
In these experiments, we only consider population agents and disabled taxes. 
The parameters of the previous experiments shown in Table \ref{tab:(hyper)parameters} are reused with a few exceptions:
First, we vary the number of resources and set $R_{d}$ to $\text{min}(1, log_2(\text{num resources}))$.
Next, we initialize all skills with a normal distribution around 1.0 with a standard deviation of 1.0 and allow agents to increase their skill. Whenever an agent \textit{gathers} or \textit{crafts}, the respective skill is increased by 0.5\%. This 0.5\% bonus shrinks as the agent approaches the maximum skill level of 5.
Additionally, we slightly simplified the action space by allowing the agents to only place trade orders for three prices: 3, 6 and 9.
Lastly, we increase the network size to 256 nodes per layer for all shared networks.

Figure \ref{fig:multi-agent-comparison} showcases the effect of the different methods in EconoJax with 4, and 12 different resources -- in which the agents have  a total of 30 and 86 discrete actions, respectively.
We can see that even in larger action spaces, different learning methods achieve similar returns. Intuitively, however, we expect individually trained policies to be more diverse \cite{johansonEmer2022}, as they only learn from their own observations and rival other agents. In turn, we would assume that this highly influences the outcome of the simulation.
Instead, our experiments indicate that centralized learning may well be capable of diverse learning diverse behavior.

Naturally, returns alone could arrive at similar levels via different policies. As such, we additionally measured the spread of the action distributions of each agent, where we binnend action types belonging to specific resource types (e.g. binning all \texttt{sell resource 1} actions). When averaging these action distributions over multiple evaluations, we again find no noticeable difference between the different multi-agent learning methods.

We note that, in order to produce more heterogeneous policies in centralized trained agents, some works enhance the observations with an agent id~\cite{rashidMono2020}. We included this method in our experiments and again found no noticeable differences in action distributions or final returns as shown in Figure \ref{fig:multi-agent-comparison}. In fact, the learning pattern almost exactly mirrors that of centralized training without agent ids. 

We believe that, given that the state contains sufficient information for specialized behavior (in the case of EconoJax likely by observing the skill levels),
centralized training should be effective, even in large action spaces.
Centralized training offers the advantage of reduced computational demands as well as allowing the population to grow dynamically without the need to heavily retrain agents. 

\section{Discussion \& Future Work} \label{sec:discussion}

\balance

EconoJax allows for fast economic simulations when compared to the AI economist. 
However, many steps can still be taken to improve the realism of the simulation.
Due to the increased performance, faster iterations during development are possible, and as such, we hope EconoJax inspires researchers to use it as a stepping stone for developing more realistic simulations.

The experiments conducted in this paper highlight emergent real-world economic behavior. However, this behavior is significantly affected by manually set parameters. In particular, the utility function of the agents has a tunable parameter, and the coins and labor obtained by each action are set manually as well. All these options heavily influence the simulation. 
As such, in addition to more functionality added to EconoJax, a necessary improvement would be to align these parameters with the real world, for example by learning them from real data. 
Nevertheless, due to the manually chosen parameters, care must be taken when building and using simulations such as EconoJax. There is a risk that individuals might unintentionally or deliberately present certain outcomes as genuine economic effects, even though the simulation was designed to produce those effects.
However, we believe that the current level of realism in EconoJax is not advanced enough to pose a significant ethical threat to society.

We experimented with different multi-agent methods and action space sizes and observed that centralized training may provide adequate diverse behavior even in action spaces of size 86. Future work may experiment with even larger experiments and provide a better metric to measure diversity. 

\section{Conclusion} \label{sec:conclusion}

This paper presented EconoJax, a multi-agent economic simulation written in JAX.
EconoJax is a simulation environment in which real-world economic behavior naturally occurs without manually programming specific policies and is largely based on the economy built in the AI economist~\cite{zhengAI2022}.
In contrast to the AI economist, EconoJax allows for fast training of large populations. Compared to their RLlib CPU-based implementation, we saw training time improvements of over 2.400 times.
We conducted experiments with population sizes of 100 and found real-world economic behavior occuring such as \textit{specialization}, the \textit{productivity-equality tradeoff}, and realistic \textit{tax schedules} and \textit{market prices}.
EconoJax can provide a stepping stone for the development of more realistic economic simulations.
We further showcased how different multi-agent learning methods did not have a significant effect on the policies produced after training on varying action space sizes.
% We hope that our results inspire researchers to build upon EconoJax for advanced economic modeling.

\begin{acks}
This work was in part performed using the compute resources from the Academic Leiden Interdisciplinary Cluster Environment (ALICE) provided by Leiden University.
This work was financially supported by Shell Information Technology International Limited and the Netherlands Enterprise Agency under the grant PPS23-3-03529461.
We would lastly like to thank Felix Kleuker for being a mathematician.
\end{acks}

%%% The following command should be issued somewhere in the first column 
%%% of the final page of your paper.

%%%%%%%%%%%%%%%%%%%%%%%%%%%%%%%%%%%%%%%%%%%%%%%%%%%%%%%%%%%%%%%%%%%%%%%%

%%% The next two lines define, first, the bibliography style to be 
%%% applied, and, second, the bibliography file to be used.

\bibliographystyle{ACM-Reference-Format} 
\bibliography{biblio}

\appendix

\newpage

\nobalance

\section{Environment Differences Compared to the AI Economist}

This section aims to summarize implementation differences between the EconoJax environment and the original AI economist environment. 

EconoJax agents do not move in a 2D world in which they have to collect resources and build houses. The `move` action has therefor been replaced by a per-resource `gather` action which allows an agent to instantly receive a resource. This additionally means that, in the base version of EconoJax, resources are not scarce. However, time is still scarce, forcing agents to specialize. Furthermore, building a house no longer requires a free spot in the world, and agents can not block each other via the building of houses.

Due to the removal of the 2D spatial world, EconoJax agents observe a flat input space, rather than a combined 2D and 1D space. Additionally, the observations of the market have been simplified to relevant statistics (highests/lowests/number of orders) instead of the full market.

The government agent's observations have also been altered. In the AI Economist, individual states about each agent were observed. In EconoJax, the government observes mean, median, and std statistics for various population metrics. This is largely because EconoJax can comfortably feature a much larger population count, and feeding observations for each would be infeasible and unrealistic from a policy-making standpoint.

EconoJax is less explicit about the type of resources available in the world. Rather than "Wood" and "Stone", there is simply $n$ amount of resources available, and crafting these into coins requires a configurable amount of resources.

Optionally, in EconoJax, agents can improve their skill in crafting (building houses) or gathering resources by performing these actions. The rate at which their skill improves is configurable.

The EconoJax resource market differs slightly in the way the price is determined for a matched trade. In the AI Economist, when two orders are matched, the price is set to whichever order was placed \textit{first}. In EconoJax, we use the price of the order that was placed \textit{last}.
This is altered because (in both environments) the lowest/highest asks/bids are evaluated for matching before other orders. If the price were to be set by the \textit{first} order, agents could exploit the market by placing only very low/high orders whenever a matching order already exists.

%%%%%%%%%%%%%%%%%%%%%%%%%%%%%%%%%%%%%%%%%%%%%%%%%%%%%%%%%%%%%%%%%%%%%%%%

\end{document}